**The Micro-Randomized Trial for Developing Digital Interventions:**

**Data Analysis Methods**

Tianchen Qian[1]

Harvard University

Michael A. Russell[2]

Pennsylvania State University

Linda M. Collins[3]

Pennsylvania State University

Predrag Klasnja[4]

University of Michigan

Stephanie T. Lanza[5]

Pennsylvania State University

Hyesun Yoo[6]

University of Michigan

Susan A. Murphy[7]

Harvard University

**Author Note**

[1]Tianchen Qian, Department of Statistics, Harvard University. Dr. Qian was supported by National Institutes of Health grants P50DA039838, R01AA023187, U01CA229437, and U54EB020404.

[2]Michael A. Russell, The Methodology Center and Department of Biobehavioral Health, The Pennsylvania State University. Dr. Russell was supported by National Institutes of Health grant P50DA039838.

[3]Linda M. Collins, The Methodology Center and Department of Human Development & Family Studies, The Pennsylvania State University. Dr. Collins was supported by National Institutes of Health grants P50DA039838, R01AA022931, P01CA180945, and R01DA040480.

[4]Predrag Klasnja, School of Information, University of Michigan, & Kaiser Permanente Washington Health Research Institute. Dr. Klasnja was supported by National Institutes of Health grants R01HL125440, U01CA229445, and R01LM013107.

[5]Stephanie T. Lanza, Edna Bennett Pierce Prevention Research Center and Department of Biobehavioral Health, The Pennsylvania State University. Dr. Lanza was supported by National Institutes of Health grant P50DA039838.

[6]Hyesun Yoo, Department of Statistics, University of Michigan. Dr. Yoo was funded by National Institutes of Health grant R01AA023187.

[7]Susan A. Murphy, Departments of Statistics and Computer Science, Harvard University. Dr. Murphy was supported by National Institutes of Health grants P50DA039838, R01AA023187, U01CA229437, UG3DE028723, and U54EB020404.

The corresponding author is Susan A. Murphy, Science Center 400 Suite, Harvard University, One Oxford Street, Cambridge, MA 02138-2901

The authors thank Amanda Applegate for helpful comments.

**Abstract**

Although there is much excitement surrounding the use of mobile and wearable technology for the purposes of delivering interventions as people go through their day-to-day lives, data analysis methods for constructing and optimizing digital interventions lag behind. Here, we elucidate data analysis methods for primary and secondary analyses of micro-randomized trials (MRTs), an experimental design to optimize digital just-in-time adaptive interventions. We provide a definition of causal "excursion" effects suitable for use in digital intervention development. We introduce the weighted and centered least-squares (WCLS) estimator which provides consistent causal excursion effect estimators for digital interventions from MRT data. We describe how the WCLS estimator along with associated test statistics can be obtained using standard statistical software such as SAS (SAS Institute Inc., 2019) or R (R Core Team, 2019). Throughout we use HeartSteps, an MRT designed to increase physical activity among sedentary individuals, to illustrate potential primary and secondary analyses.

# The Micro-Randomized Trial for Developing Digital Interventions:
# Data Analysis Methods

Mobile technologies—including tablets, smartphones, and wearable sensors—have become ubiquitous in daily life. Because of their ability to engage users "in the moment," they provide unprecedented opportunity to deliver interventions at the times and in the contexts when individuals are most likely to benefit. Just-in-time adaptive interventions (JITAIs; Nahum-Shani et al., 2018) are intervention designs that take advantage of these opportunities in order to intensively adapt and deliver interventions to each individual in the flow of their life. JITAI design involves the construction of decision rules that pinpoint the best intervention option for the individual's current context. However, research methods useful in optimizing JITAI decision rules lag behind technology's current capabilities to reach individuals, with the result that scientific knowledge is lacking about what intervention content should be delivered when, how often, and in which context, without overburdening the individual. The micro-randomized trial (MRT) is an experimental trial for obtaining this knowledge. In an MRT, intervention component options are randomized to each participant many times during the trial (see the companion article, Walton et al. (under review), for more details and examples of MRTs). As discussed in Walton et al. (under review), the MRT was developed for use in constructing and optimizing JITAI decision rules in digital interventions (Klasnja et al., 2015; Nahum-Shani et al., 2018).

A natural primary analysis of data from an MRT would focus on whether there is a marginal causal effect of an intervention component. Possible secondary analyses might include moderation analyses aimed at informing JITAI development. As will be shown below, the meaning of marginal and moderation causal effects requires careful thought. The purpose of this article is to describe these causal effects and discuss ways to analyze the data from an MRT in order to optimize a JITAI.

We begin by providing a brief review of the MRT, along with an overview of the HeartSteps MRT (Klasnja et al., 2018) for use in clarifying ideas. We then offer precise definitions of marginal and moderation effects via the concept of causal "excursion" effects, using the potential outcomes framework (Robins, 1986; Rubin, 1978). Next we review and explain the weighted and centered least-squares (WCLS) estimator (Boruvka et al., 2018), which provides estimators and test statistics for conducting primary and secondary analyses using data from MRTs. Data from the HeartSteps MRT is used to illustrate these analyses.

**Brief Review of MRTs**

MRTs are conducted with the goal of providing intensive longitudinal data that can be used to develop one or more JITAI components. Each component is associated with different intervention options that might be provided at any of multiple *decision points* during an individual's day-to-day life. For example, one of the components examined in the HeartSteps MRT was the activity suggestions component, for which the intervention options were a contextually tailored activity suggestion[1] or no suggestion. The activity suggestions component was randomly assigned at 5 decision points per day.

In an MRT, each participant is repeatedly randomized to different options of an intervention component, with known probability at each decision point. This repeated, intensive randomization means that over the course of an MRT, a participant may be randomized hundreds or even thousands of times. As discussed in the companion article (Walton et al., under review), intervention components are often designed to have greatest impact on a near-term *proximal outcome*. The proximal outcome is observed after each randomization. Prior to each decision point, observations of context such as sensor data (e.g., heart rate, step count, weather, current location), as well as self-report data (perceptions of loneliness, perceptions of social isolation, enjoyment of physical activity), may be observed. As discussed in the companion article (Walton et al., under review), observations of context may be used to inform the content of a message in an intervention option and to define availability conditions (defined in the next paragraph). As will be shown below, some observations might be collected to serve as controls in data analyses or to inform moderation analyses.

---

[1] The contextually tailored activity suggestion at each time is delivered in one of two forms (with equal probability): either a suggestion with a walking activity that took 2-5 minutes to complete or an anti-sedentary suggestion that instructs brief movements such as to stand up and roll one's arms; see Walton et al. (under review). For expositional simplicity we group all activity suggestions together for most parts of this article.

As an individual goes through everyday life, there may be times when only the intervention option of "no treatment" is appropriate. This is formalized in the notion of *availability*. For example, often the delivery of the intervention involves an audible and visual cue. If sensors on the phone and/or wearables indicate that the individual might be operating a moving vehicle at a decision point, then to avoid potentially dangerous distractions the individual might be deemed unavailable for an intervention. Individuals also may be deemed unavailable if an intervention was delivered within the prior $x$ minutes (to reduce burden). At times of unavailability, the only appropriate intervention is the "no treatment" option. Availability is one of the time-varying variables observed prior to each decision point.

**HeartSteps**

The HeartSteps study involves a 42-day MRT designed to inform the optimization of the HeartSteps digital intervention to increase physical activity (Klasnja et al., 2018; Walton et al., under review). HeartSteps combines a wristband activity tracker that monitors participants' steps throughout the day in concert with a mobile phone application. Two intervention components were investigated in the HeartSteps MRT: a planning support component and the contextually tailored activity suggestions component described above. The planning support component consisted of support for planning in the evening for activity on the next day. The intervention options were *planning* and *no planning*[2]. For simplicity, much of the exposition below focuses on the activity suggestions component.

Activity suggestions were randomized at 5 decision points each day: morning commute, lunch time, mid-afternoon, evening commute, and after dinner. The exact time of day of the five decision points was specified by each participant at the beginning of the study and could differ between participants. At each decision point, if the participant was available, the probability of delivering a contextually tailored activity suggestion (as opposed to no suggestion) was .6. The proximal outcome for the activity suggestion component was the step count of the participant in the 30-minute window following a decision point.

In the case of the activity suggestions component, as mentioned above, a participant was deemed unavailable at a decision point either if their smart device indicated that the participant might be driving, or if the participant had turned off intervention delivery (participants could turn off delivery over the subsequent 1, 2, 4 or 8 hours). Furthermore, because the content involved suggestions for new activities, a participant was deemed unavailable if they were currently walking or running or they had just finished an activity bout in the previous 90 seconds prior to the decision point.

A natural primary research question motivating the HeartSteps MRT is whether there is a causal effect of delivering an activity suggestion versus not delivering any suggestion on the proximal outcome—i.e., the subsequent 30-minute step count. This question can be addressed by estimating a causal effect that is conceptually similar to the main effect of a factor (Collins et al., 2009) because, as is explained below, it is marginal over the other intervention components. Additional important research questions in the development of the HeartSteps activity suggestion component concern the potential effects of habituation (Rankin et al., 2009) and/or treatment burden (Clawson et al., 2015; Eysenbach, 2005; Ho & Intille, 2005; Klasnja et al., 2008; Shaw et al., 2013; Yardley et al., 2016). If individuals habituate to the activity suggestions or find the intervention burdensome, the causal effect would be expected to deteriorate over time. Thus, a natural secondary or exploratory analysis is to assess whether day in study moderates the effect of delivering an activity suggestion. Two additional examples of secondary research questions are whether the effect of delivering an activity suggestion depends on the individual's current location and whether the effect of the activity suggestion depends on whether the individual was prompted to plan an activity for that day. All of these research questions as we have stated them are imprecise about what is meant by a causal effect. In the following section we offer a more precise definition of a causal effect as estimated in an MRT.

---

[2] An individual is sent a planning prompt with probability .5 every day. If a planning prompt is sent, it is delivered in one of two forms (with equal probability): structured planning (where the individual is prompted to select a plan from a list of their own past activity plans) or unstructured planning (where the individual is prompted to type their plan into a text box). For expositional simplicity we group both forms of planning together in this paper.

## The Causal Excursion Effect

In this section, we define causal effects using the potential outcomes framework (Robins, 1986, 1987; Rubin, 1978). For expositional clarity, throughout the paper we consider the setting in which there are only two intervention options, denoted by treatment 1 and treatment 0. For the activity suggestions component in the HeartSteps MRT, this would be delivering activity suggestion (treatment 1) and not delivering activity suggestion (treatment 0). First, we briefly review the definition of a causal effect using a hypothetical setting with a single time point treatment. Then we define the causal excursion effect of a time-varying intervention component on a time-varying outcome. Throughout, upper case letters denote random variables and lower case letters denote particular values of the random variables.

In the classical potential outcomes framework, where there is only a single time point for possible treatment (see review by Rubin (2005)), the ideal but usually unattainable goal is to determine the individual-level causal effect, or the difference between the outcome under treatment 1 [denoted by $Y(1)$] and the outcome under treatment 0 [denoted by $Y(0)$] at the same time for each individual. Consider the first decision point in the HeartSteps MRT. At this decision point individuals are randomly assigned to receive an activity suggestion or no suggestion. The step count in the 30-minute window following this decision point is the outcome. For each individual, the treatment effect at this decision point is the difference between (a) the 30-minute step count had treatment been assigned to the individual ($Y(1)$) and (b) the 30-minute step count had the treatment not been assigned to the individual ($Y(0)$). $Y(1)$ and $Y(0)$ are called *potential outcomes*, because in reality only one of the potential outcomes can be observed on each individual, as both treatment and no treatment cannot be assigned to an individual at the same time—this is the "fundamental problem of causal inference" (Holland, 1986).

Let $A$ denote the treatment assignment ($A = 1$ if treatment 1; $A = 0$ if treatment 0). Only $Y = AY(1) + (1 - A)Y(0)$ is observed.[3] A widely adopted solution to this problem is to estimate a marginal causal effect, or to estimate an effect closer to the individual-level causal effect, the causal effect conditional on a pre-treatment variable $S$; that is, the difference between the expected outcome had everyone with $S = s$ received the treatment ($E[Y(1)|S = s]$) and the expected outcome had everyone with $S = s$ *not* received the treatment ($E[Y(0)|S = s]$). In the example for the first decision point in the HeartSteps MRT, $S$ might be the individual's current location (home, work or other), current weather, gender, and baseline activity level. An interesting scientific question would be whether the value of $S$ modifies the treatment effect. Inference for the difference, $E[Y(1)|S = s] - E[Y(0)|S = s]$, answers this question.

---

[3] This equality holds under the consistency assumption often made in causal inference literature, which essentially requires that there are no two "versions" of the same treatment. In the example of activity suggestions, in order to properly define "delivering an activity suggestion" as treatment 1 and "not delivering an activity suggestion" as treatment 0, one would consider various framings and various contents of the suggestions as a "compound treatment." However, if one wishes to distinguish between the effect of different versions of the suggestions in the analysis, then instead it would be necessary to define the treatment to have multiple levels.

If $A$ is randomized, then the above difference in terms of potential outcomes can be written in terms of expectations with respect to the distribution of the observations $(S, A, Y)$. In particular, if treatment is randomly assigned, the causal effect, $E[Y(1)|S = s] - E[Y(0)|S = s]$, is equal to $E[Y|A = 1, S = s] - E[Y|A = 0, S = s]$ (see Rubin (2005)).

To define the causal excursion effect of a time-varying intervention component on a time-varying outcome, notation is needed to accommodate time. Consider the HeartSteps MRT, in which one goal is to estimate the effect of the activity suggestion on the subsequent 30-minute step count. Recall that the HeartSteps MRT is a 42-day study and there were 5 decision points per day for the activity suggestion component; thus, there are $T = 210$ decision points overall. Let $X_t$ represent the data available from decision point $t - 1$ up to and including decision point $t$.[4] $X_t$ contains the availability indicator, $I_t$, with $I_t = 1$ meaning that the individual is available at decision point $t$ and $I_t = 0$ otherwise. In the HeartSteps example, $X_t$ also includes current weather, location, time of day, 30-minute step count prior to decision point $t$, and whether planning support was provided on the prior evening. Let $A_t$ represent the treatment indicator at decision point $t$, where $A_t = 1$ means treatment is delivered and $A_t = 0$ means treatment is not delivered. We use an overbar $\bar{A}_t$ to denote all prior and present treatments $\bar{A}_t = (A_1, \ldots, A_t)$.[5] Let $Y_{t+1}$ represent the proximal outcome—here, the number of steps in the 30 minutes after decision point $t$. Denote by $H_t$ the individual's history of data observed up to decision point $t$: $H_t = (X_1, A_1, Y_2, \ldots, X_{t-1}, A_{t-1}, Y_t, X_t)$. $H_t$ includes potential moderators and control variables, some of which may be composites made up of other variables in $H_t$. We denote potential moderators by $S_t$. Recall that potential moderators of the effect of the activity suggestions in HeartSteps, $S_t$, include number of days in treatment, the individual's current location (home, work or other), current weather, whether planning was prompted the prior evening, number of activity suggestions delivered on the prior day, gender, and baseline activity level. As in the single time point setting, the inclusion of potential moderators, $S_t$, means that the desired causal excursion effect is conditional on these variables. Further discussion regarding this point is included towards the end of this section after the formal definition of the causal excursion effect.

---

[4] For simplicity, we omit the subscript $i$ for the $i^{th}$ individual in $X_{it}$ and in all other variables unless necessary.

To define the causal excursion effect, we use an extension of the potential outcomes framework to the setting of intensive longitudinal data (Robins, 1986, 1987). Recall that lower case letters such as $a_t$ represent particular values of a random variable, here a possible value of the treatment $A_t$. Recall we use the overbar to represent present and past values, that is, $\bar{a}_t = \{a_1, a_2, \ldots, a_t\}$. The potential outcomes for $Y_{t+1}, X_t, H_t, S_t$ are $Y_{t+1}(\bar{a}_t), X_t(\bar{a}_{t-1}), H_t(\bar{a}_{t-1}), S_t(\bar{a}_{t-1})$, respectively. For example, $Y_{t+1}(\bar{a}_t)$ is the 30-minute step count outcome after decision point $t$ that would have been observed if the individual had been assigned treatment sequence $\bar{A}_t = \bar{a}_t$. (For binary treatments, there could be $2^t$ different potential outcomes, $Y_{t+1}(\bar{a}_t)$.) This notation encodes the reality that an individual's 30-minute step count outcome after decision point $t$ may be impacted by all prior treatments, as well as the current treatment. Note that unlike the potential proximal outcome $Y_{t+1}(\bar{a}_t)$, potential outcomes for $X_t, S_t$, as well as availability, $I_t$, are indexed *only* by treatments, $\bar{a}_{t-1}$, prior to decision point $t$—namely, $X_t(\bar{a}_{t-1})$, $S_t(\bar{a}_{t-1})$, and $I_t(\bar{a}_{t-1})$. This is because $X_t$, $S_t$, and $I_t$ occur prior to treatment at $t$.

---

[5] Note that the overbar in $\bar{A}_t$ is not an abbreviation for the average; rather, it stands for the entire vector of treatment assignment $(A_1, \ldots, A_t)$ and similarly for $\bar{a}_t$. This notation is common in causal inference literature (e.g., Robins, 1986, 1987).

The *causal excursion effect* of activity suggestions at decision point $t$ on subsequent 30-minute step count for available individuals with $S_t = s$ is defined as (Boruvka et al., 2018; Liao et al., 2016)

$$\beta(t,s) = E[Y_{t+1}(\bar{A}_{t-1}, 1) - Y_{t+1}(\bar{A}_{t-1}, 0) \mid I_t(\bar{A}_{t-1}) = 1, S_t(\bar{A}_{t-1}) = s]. \quad (1)$$

This formula contains the following information.

1. The effect, $\beta(t,s)$, is *causal* because it is the expected value of the contrast in step counts in the 30 minutes following a decision point $t$ if the treatment were delivered at $t$ [potential outcome $Y_{t+1}(\bar{A}_{t-1}, 1)$] versus if treatment were not delivered at $t$ [potential outcome $Y_{t+1}(\bar{A}_{t-1}, 0)$]; that is, $Y_{t+1}(\bar{A}_{t-1}, 1) - Y_{t+1}(\bar{A}_{t-1}, 0)$.

2. The effect, $\beta(t,s)$, is *conditional*. This effect is only among individuals who are available ($I_t(\bar{A}_{t-1}) = 1$) and for whom the potential moderators take on the value of ($S_t(\bar{A}_{t-1}) = s$).

3. The effect, $\beta(t,s)$, is *marginal*. In HeartSteps, the effect of the activity suggestions component at decision point $t$ is marginal over potential moderators not contained in $s$, the effects of interventions from prior decision points, and the planning support component. In fact, $\beta(t,s)$ is replaced by $\beta(t)$ if, for example, estimation of the marginal effect of delivering an activity suggestion compared to no activity suggestion is desired; in this case $S_t(\bar{A}_{t-1})$ is omitted from (1).

4. The effect, $\beta(t,s)$, is an *excursion* from the "treatment schedule" prior to $t$. In an MRT the treatment schedule prior to $t$ is a set of probabilistic decision rules for treatment assignment at all decision points from the beginning of the intervention up to the previous decision point; that is, for assignment of $A_1, \ldots, A_{t-1}$. In the case of an MRT, the treatment schedule will always involve some randomization, but may include non-random assignment as well. For example, in the HeartSteps MRT the treatment schedule included, at five decision points per day, the following: if available deliver an activity suggestion with probability .6 and no suggestion with probability .4; if not available do not deliver an activity suggestion. Suppose the HeartSteps intervention included a component that was not examined in the MRT—for example, a brief motivational video sent to all participants every Monday morning at 8 am. Although there is no experimentation on this component, it would be included in the treatment schedule.

The causal excursion effect concerns what would happen if an individual followed the current treatment schedule up to time $t-1$ and then deviated from the schedule to assign treatment 1 at decision point $t$, versus deviated from the schedule to assign treatment 0 at decision point $t$. In other words, the definition of $\beta(t,s)$ implicitly depends on the schedule for assigning $A_1, \ldots, A_{t-1}$. Technically this excursion can be seen from (1), in that the expectation, $E$, is marginal over all prior treatments not contained in $S_t$. For example, if $S_t$ contains only current weather, then the excursion effect is marginal over all the variables other than current weather, including the schedule for assigning all of the prior treatments, $A_1, \ldots, A_{t-1}$, as well as all prior treatments for other components such as the planning component.

To understand the excursion effect better, consider two very different treatment schedules. In the first schedule, the treatment is provided on average once every other day; in the second schedule, the treatment is provided on average 4 times per day. Excursions from these two rather different schedules could result in very different effects, $\beta(t, s)$. Indeed, in the latter schedule individuals may experience a great deal of burden and disengage with the result that $\beta(t, s)$ would be close to 0, whereas in the former schedule individuals may still be very engaged, resulting in a larger $\beta(t, s)$. In other words, the definition of the causal excursion effect, $\beta(t, s)$, is dependent on the schedule for treatment assignment; this is different from the types of effects typically discussed in the causal inference literature (e.g., Robins (1994); Robins, Hernán, & Brumback (2000)).

A primary hypothesis test might focus on inference about the marginal excursion effect, that is, (1) with $S_t(\bar{A}_{t-1})$ equal to an empty set. A secondary analysis might consider treatment effect moderation by including in $S_t(\bar{A}_{t-1})$ potential moderators, such as location, or number of days in the intervention. Note that $S_t(\bar{A}_{t-1})$ does not need to include all true moderators for (1) to be a scientifically meaningful causal effect; instead, it is appropriate to choose any (or no) $S_t(\bar{A}_{t-1})$, provided that (1) is interpreted as the causal excursion effect marginal over all variables in $H_t(\bar{A}_{t-1})$ that are not included in $S_t(\bar{A}_{t-1})$.

Under the assumptions that (i) the treatment is sequentially randomized (which is guaranteed by MRT design) and (ii) the treatment delivered to one individual does not affect another individual's outcome[6], the causal excursion effect $\beta(t, s)$ in (1) can be written in terms of expectations over the distribution of the data as

$$\beta(t,s) = E[E(Y_{t+1} \mid A_t = 1, H_t, I_t = 1) - E(Y_{t+1} \mid A_t = 0, H_t, I_t = 1) \mid I_t = 1, S_t = s]. \quad (2)$$

---

[6] This assumption is called "non-interference" in causal inference. If there are social network components in the digital intervention, this assumption may be violated and an extension of the potential outcomes framework to incorporate interference is needed (Hong & Raudenbush, 2006; Hudgens & Halloran, 2008).

Thus the excursion effect $\beta(t,s)$ can be estimated using the observed MRT data.

**A Primary Research Question for HeartSteps**

As discussed above, a natural primary research question that can be addressed in the HeartSteps MRT is whether there is a marginal causal excursion effect of delivering an activity suggestion on the subsequent 30-minute step count of the user, compared to not delivering any message. To operationalize this marginal causal excursion effect, let $S_t$ be an empty set in (2) so that

$$\beta(t) = E[E(Y_{t+1} \mid A_t = 1, H_t, I_t = 1) - E(Y_{t+1} \mid A_t = 0, H_t, I_t = 1) \mid I_t = 1]. \quad (3)$$

The outer expectation on the right-hand side in (3) represents an average across all possible values of $H_t$ across individuals. For example, $\beta(t)$ is averaged over weather on that day and on previous days, over previous treatment assignment of the activity suggestions (i.e., $A_s$ for $s < t$), and also over previous assignment of the planning support prompts. The marginal excursion effect is the average of $\beta(t)$ over $t$ with weights equal to $E[I_t]$ to obtain

$$\beta_0 = \frac{\sum_{t=1}^{T} E[I_t]\beta(t)}{\sum_{t=1}^{T} E[I_t]}. \quad (4)$$

Here $E[I_t]$ denotes the probability of an individual being available at decision point $t$. Thus $\beta_0$ is a weighted average of the marginal effects, $\beta(t)$, in which the weights are the availability probabilities. In the section "Analysis Using Data from HeartSteps MRT," we will conduct inference about a variety of excursion effects including this marginal excursion effect, $\beta_0$ (Klasnja et al., 2018).

**A Selection of Secondary Research Questions for HeartSteps**

Consider secondary research questions that concern moderation of the causal excursion effect by a non-empty $S_t$. For example, one question might be whether the causal excursion effect deteriorates with day under treatment. In this case $S_t$ would include $\text{day}_t$, the number of days in treatment prior to the decision point $t$. Suppose the causal excursion effect can be represented by the linear model:

$$\beta(t, \text{day}_t) = E[E(Y_{t+1} \mid A_t = 1, H_t) - E(Y_{t+1} \mid A_t = 0, H_t) \mid I_t = 1, day_t] = \beta_0 + \beta_1 \text{day}_t.$$

Note that $\text{day}_t = 0$ for all decision points $t$ on the first day of treatment; thus $\beta_0$ represents the causal excursion effect on the first day; $\beta_1$ represents the change in the causal excursion effect with each additional day.

Other secondary research questions concern whether there is effect moderation by other time-varying observations, such as the current location of the user, or by another intervention component being examined in the MRT. Consider the planning support component in the HeartSteps MRT. Let $S_t$ denote the indicator of whether a planning support prompt was delivered on the evening prior to decision point $t$ ($S_t = 1$ if delivered, $S_t = 0$ if not). A model that includes effect moderation by a planning support prompt on the previous evening can be expressed as

$$\beta(t, A_{t-1}) = E[E(Y_{t+1} \mid A_t = 1, H_t) - E(Y_{t+1} \mid A_t = 0, H_t) \mid I_t = 1, S_t] = \beta_0 + \beta_1 S_t.$$

Here $\beta_0$ represents the causal excursion effect when the individual did not receive planning support on the prior evening, and $\beta_0 + \beta_1$ represents the causal excursion effect when the individual received planning support on the prior evening. See the section "Analysis Using Data from HeartSteps MRT" for results of the secondary analyses concerning these questions.

**Methods for Estimating Causal Excursion Effects from MRT Data**

We present a WCLS estimator, developed by Boruvka et al. (2018), that provides a consistent estimator for the causal excursion effect, $\beta(t, s)$. Here for clarity we provide an overview of the estimation method that can be used when the randomization probabilities are constant, as is the case in HeartSteps. Recall that in HeartSteps a primary analysis might be an assessment of the marginal causal excursion effect of the activity suggestions on the subsequent 30-minute step count. We use the superscript $T$ to denote the transpose of a vector or a matrix.

Suppose the causal excursion effect is linear: $\beta(t, s) = s^T \beta$ with $\beta(t, s)$ defined in (2), and the goal is to make inference about $\beta$. Note that the model for $\beta(t, s)$ characterizes only the treatment effect (as a contrast between the two treatments with the proximal outcomes as dependent variables). The WCLS requires a *working model* for the conditional mean of $Y_{t+1}$ given no treatment at decision point $t$ and history $H_t$ [i.e., $E(Y_{t+1}|A_t = 0, I_t = 1, H_t)$]. One working model might be $Z_t^T \alpha$, where $Z_t$ is a vector of summaries of the observations made prior to decision point $t$ (i.e., summaries constructed from $H_t$). These summaries are often called control variables. For example, in HeartSteps a natural control variable is the step count in the 30 minutes prior to the decision point. This prior 30-minute step count variable is likely strongly correlated with the proximal outcome, the 30-minute step count following the decision point. As will be seen shortly, *consistency of the WCLS estimator for $\beta$ does not require $Z_t^T \alpha$ to be a correct model for $E(Y_{t+1}|A_t = 0, I_t = 1, H_t)$*. The role of $Z_t^T \alpha$ is to reduce noise in the analysis: Inclusion of prognostic control variables (those variables in $H_t$ that are correlated with $Y_{t+1}$) in $Z_t$ will usually reduce the variance of the estimator of $\beta$. A simulation study that illustrates this point is presented in Appendix C. In summary, the inclusion of the step count in the 30 minutes prior to the decision point serves to reduce variance and increase the power to detect a nonzero excursion effect, $\beta$.

The WCLS estimator for $\beta$ is calculated as follows. Suppose $(\hat{\alpha}, \hat{\beta})$ is the $(\alpha, \beta)$ value that solves the following estimating equation (Diggle et al., 2002)

$$\sum_{i=1}^{n}\sum_{t=1}^{T} I_{it}\left[Y_{i,t+1} - Z_{it}^T\alpha - (A_{it} - p)S_{it}^T\beta\right]\begin{bmatrix} Z_{it} \\ (A_{it} - p)S_{it} \end{bmatrix} = 0, \quad (5)$$

where $0 < p < 1$ is the randomization probability and $i$ is the index for the $i$-th participant. Note that the indicator $A_{it}$ is centered by subtracting $p$. Recall that in HeartSteps, at each decision point there is .6 probability to deliver an activity suggestion (if the participant is available), so $p = .6$. The $\hat{\beta}$ that solves (5) is the WCLS estimator for $\beta$.

**Remarks**.

1. WCLS does *not* assume a model for the proximal outcome such as $Y_{t+1} \sim Z_t^T\alpha + (A_{it} - p)S_t^T\beta$. The primary assumption (Boruvka et al., 2018) that ensures that $\hat{\beta}$ is consistent is

$$E[E(Y_{t+1} \mid A_t = 1, H_t) - E(Y_{t+1} \mid A_t = 0, H_t)|I_t = 1, S_t] = S_t^T\beta; \quad (6)$$

that is, $\beta(t, s) = s^T\beta$ is a correct model for the causal excursion effect conditional on $I_t = 1, S_t = s$.

2. The centering of the treatment indicator, $(A_{it} - p)$, in (5) creates orthogonality between the columns of the design matrix for the causal excursion effect (i.e., $(A_{it} - p)S_t$) and the columns of the design matrix involved in the working model, $Z_t^T\alpha$, for $E(Y_{t+1}|A_t = 0, I_t = 1, H_t)$ (i.e., $Z_t$). This centering provides robustness in the estimation of $\beta$; in particular, robustness against a mis-specified working model for $E(Y_{t+1}|A_t = 0, I_t = 1, H_t)$. In other words, the data analyst can use a possibly incorrect working model $Z_t^T\alpha$, and $\hat{\beta}$ will still be a consistent estimator of $\beta$. In digital interventions this robustness property is of practical importance because vast amounts of data (i.e., high-dimensional $H_t$) on the participant have usually been collected prior to decision point $t$. As a result, it is virtually impossible to correctly model $E(Y_{t+1}|A_t = 0, I_t = 1, H_t)$. For example, in HeartSteps there are 210 decision points (210 = 42 days × 5 times/day) for each participant; $H_t$ can include the outcome, treatment, and covariates from all the past $t - 1$ decision points, which means hundreds of variables at a later decision point $t$. In addition, $E(Y_{t+1}|A_t = 0, I_t = 1, H_t)$ may depend on variables in $H_t$ in a nonlinear way, which adds to the difficulty of correctly modeling $E(Y_{t+1}|A_t = 0, I_t = 1, H_t)$ and thus makes the robustness property desirable.

3. While the choice of $Z_t$ doesn't affect the consistency of $\hat{\beta}$, a better working model for $E(Y_{t+1}|A_t = 0, I_t = 1, H_t)$ has the potential to decrease the variance of $\hat{\beta}$. Because $H_t$ is usually high-dimensional, choosing $Z_t$ can be done by hand-picking a subset of $H_t$ (e.g., those covariates and outcomes at recent decision points). As discussed above, in HeartSteps a natural control variable is the step count in the 30 minutes prior to the decision point, as this variable is likely highly correlated with the proximal outcome of step count in the 30 minutes after the decision point. In Appendix C we illustrate through a simulation study how inclusion of control variables that are correlated with the proximal outcome in $Z_t$ can reduce the variance of $\hat{\beta}$.

4. Although software based on the estimating equation (5) also outputs an estimator $\hat{\alpha}$ and its standard error, we recommend not interpreting them, unless it is safe to assume that $Z_t^T \alpha$ is a correct model for $E(Y_{t+1}|A_t = 0, I_t = 1, H_t)$.

For simplicity in presentation of the estimator, so far we have considered the setting where the randomization probability, $p$, is constant. There are also practical settings where the randomization probability may change over time; for example, in the stratified micro-randomized trial, different micro-randomization probabilities are used depending on a time-varying variable such as prediction of risk. For example, a higher randomization probability may be used when the individual is categorized as high-risk, and a lower randomization probability may be used when the individual is categorized as low-risk. The rationale for such risk stratification is to ensure that there are adequate numbers of treatments delivered both at risk times and at non-risk times. See Dempsey, Liao, Kumar, & Murphy (2019) for details. The WCLS estimator presented above can be generalized to this setting (Boruvka et al., 2018); see Appendix B for a general WCLS estimator that allows randomization probability to depend on the individual's history, $H_t$.

**Estimating the WCLS $\hat{\beta}$ Using Standard Statistical Software**

When the randomization probabilities are constant, standard statistical software that implements GEE (Liang & Zeger, 1986), such as SAS (SAS Institute Inc., 2019), Stata (StataCorp, 2019), and SPSS (IBM Corp., 2019), can be "tricked" into providing the WCLS estimator $\hat{\beta}$ and its standard error. Consider SAS PROC GENMOD (SAS Institute Inc., 2019) and suppose the assumed causal excursion effect model is (6) and the working model for $E(Y_{t+1}|A_t = 0, I_t = 1, H_t)$ is $Z_t^T \alpha$. Then the WCLS estimator $\hat{\beta}$ and its standard error can be obtained by the following steps: (i) incorporate $I_t$ as the "prior weights," (ii) choose a working independence correlation structure, and (iii) fit GEE with dependent variable $Y_{t+1}$ and independent variables $Z_t$ and $(A_t - p)S_t$. Then the estimated coefficient for $(A_t - p)S_t$ is the WCLS estimate $\hat{\beta}$. Note that in choosing the control variables, $Z_t$ needs to contain at least $S_t$.

This approach does not actually fit a GEE model for the conditional mean of $Y_{t+1}$. Instead, this estimation method merely uses the GEE software as a means to output the WCLS estimator. Technically, this can be done because the estimating equation of a GEE with the above specification is algebraically equivalent to (5), the estimating equation of WCLS. To obtain appropriate standard errors for the estimator $\hat{\beta}$ through the above GEE fit, one needs to use the robust standard error [SAS calls this the "empirical standard error" (SAS Institute Inc., 2019)]. The robust standard error accounts for the correlations among the proximal outcomes, $Y_2, Y_3, \ldots, Y_{T+1}$. When the sample size is small (e.g., $n < 50$), we recommend use of further small sample corrections for both the standard error and the degrees of freedom in the critical value for constructing confidence intervals (Boruvka et al., 2018). R code (R Core Team, 2019) for the implementation with the small sample correction is available at

https://github.com/StatisticalReinforcementLearningLab/HeartstepsV1Code/blob/master/xgeepack.R.

### Analysis Using Data from HeartSteps MRT

Recall that HeartSteps is a 6-week MRT for optimizing JITAI components of a digital intervention to promote physical activity with 37 sedentary participants (Klasnja et al., 2018). In the illustrative analysis below, we focus on the activity suggestion component, which was randomized at 5 decision points each day. At each decision point, if the participant was available, an activity suggestion was delivered with randomization probability .6. We first address the primary research question by estimating the marginal excursion effect of an activity suggestion versus no suggestion using data from the HeartSteps MRT. The primary analysis for HeartSteps is published in Klasnja et al., (2018); for completeness we include this analysis as well as results of additional secondary analyses. As discussed before, secondary research questions might include how the excursion effect changes over time, whether this effect is moderated by current location, and whether this effect is moderated by delivery of a planning support prompt on the evening prior to the decision point. All analyses are conducted using the R programming language (R Core Team, 2019). We use the following variables in the analysis:

- $Y_{t+1}$: log-transformed 30-minute step count following decision point $t$. This is the proximal outcome of interest.

- $A_t$: indicator of whether an activity suggestion is delivered at decision point $t$. The randomization probability is .6 at available decision points.

- $X_{t,1}$: log-transformed 30-minute step count preceding decision point $t$. Because this variable is expected to be correlated with $Y_{t+1}$, we will include $X_{t,1}$ as a control variable in the analysis to reduce noise.

- $X_{t,2}$: day in the study, coded as 0, 1, 2, …, 41.

- $X_{t,3}$: participant's location at decision point $t$; coded as 1 if at home or at work, and 0 if at any other location.

- $X_{t,4}$: indicator of a planning support prompt delivered on evening prior to decision point $t$; coded as 1 if delivered and 0 if not.

- $I_t$: availability status at decision point $t$. Recall that randomization can occur only if the participant is available.

Step count data are highly skewed; the log-transformation is used to make its distribution more symmetric (and we added .5 to the step count before taking log to avoid log(0)). Although the consistency of the WCLS estimator does not require $Y_{t+1}$ to be symmetrically distributed, symmetry improves the accuracy of the approximation to the distribution of the test statistic in small samples.

**Question 1: On average, is there an effect of delivering an activity suggestion on subsequent 30-minute step count, compared to no suggestion?**

As discussed above a natural primary research question concerns the excursion effect marginal over all decision points and all covariates. We address this question using the WCLS estimator with $S_t$ equal to the empty set, $\beta(t,s) = \beta_0$, working model $\alpha_0 + \alpha_1 X_{t,1}$, and weight $I_t$. Table 1 lists the results. The causal effect of delivering an activity suggestion versus no suggestion on the log-transformed subsequent 30-min step count, averaged over all decision points and all covariates, is $\hat{\beta}_0 = 0.131$ ($p = 0.060$, 95% CI = -0.006 to 0.268). This corresponds roughly to a 14% ($= e^{0.131} - 1$) increase in the average 30-minute step count (on its original scale), comparing decision points when an activity suggestion was sent with decision points when an activity suggestion was not sent.

**Question 2: Does the effect of activity suggestions change with each additional day in the study?**

This question is motivated by the hypothesis that the longer a person participates in the study, the more they may habituate to the suggestions or become overburdened, leading them to become less responsive. We address this question via the WCLS estimator with $S_t = X_{t,2}$ (day in study), $\beta(t,s) = \beta_0 + \beta_1 X_{t,2}$, working model $\alpha_0 + \alpha_1 X_{t,1} + \alpha_2 X_{t,2}$, and weight $I_t$. $X_{t,2}$ is included in $S_t$ to assess the effect moderation by $X_{t,2}$, day in the study. Because $X_{t,2}$ is coded to start from 0, $\beta_0$ represents the causal excursion effect on the first day. Table 2 lists the results. There is a significant interaction between the activity suggestion and day in the study: the causal effect of the activity suggestion changes by $\hat{\beta}_1 = -0.018$ with each additional day in the study ($p = 0.005$, 95% CI = -0.031 to -0.006). Combining this with $\hat{\beta}_0 = 0.507$, the analysis indicates that sending an activity suggestion results in about 66% ($= e^{0.507} - 1$) increase in the 30-minute step count on the first day of the study, about 16% ($= e^{0.507-0.018\times 20} - 1$) increase on the 21st day of the study, and about 21% ($= 1 - e^{0.507-0.018\times 41}$) decrease on the 42nd day of the study. A sensitivity analysis to the linearity assumption (that the causal excursion effect changes linearly by day in the study) is provided in Appendix D.

**Question 3: Does the effect of delivering *each type of* activity suggestion versus no suggestion depend on the individual's current location (home/work, or other)?**

The activity suggestion involves suggestions for new physical activities; therefore, it is of interest to examine whether its effect depends on the individual's location, which might be a proxy for interruptibility. If an activity suggestion is sent, then ½ of the time the suggestion is a walking suggestion (instructing a walking activity that took 2-5 minutes to complete) and the remaining ½ of the time the activity suggestion is an anti-sedentary suggestion (instructing brief movements, such as stretching one's arms). The investigators conjectured that effect moderation by location may differ between walking suggestions and anti-sedentary suggestions. Therefore, here we assess whether the effect of delivering each type of activity suggestion versus no suggestion is modified by the individual's current location (home/work or other). We address this question by using the WCLS estimator with $S_t = X_{t,3}$ (indicator of being at home or work), working model $\alpha_0 + \alpha_1 X_{t,1} + \alpha_2 X_{t,3}$, and weight $I_t$. $X_{t,3}$ is included in $S_t$ to assess the effect moderation by $X_{t,3}$, location of the individual. Because here we have two treatment indicators (indicator of whether a walking suggestion is delivered, and indicator of whether an anti-sedentary suggestion is delivered), the causal excursion effect for the walking suggestion is modeled as $\beta_0 + \beta_1 X_{t,3}$, and the causal excursion effect for the anti-sedentary suggestion is modeled as $\beta_2 + \beta_3 X_{t,3}$. Table 3 lists the result. The causal excursion effect moderation by location (home/work or other) is statistically significant for walking suggestions ($\hat{\beta}_1 = 0.377, p$

= 0.049, 95% CI = 0.001 to 0.753). The effect moderation is not statistically significant for anti-sedentary suggestions ($\hat{\beta}_3 = -0.142$, $p = 0.472$, 95% CI = -0.540 to 0.256).

**Question 4: Does the effect of activity suggestions depend on whether planning support was delivered on the previous evening?**

Whether planning support was delivered on the previous evening may impact the effectiveness of the activity suggestion. To assess this moderation effect, we use the WCLS estimator with $S_t = X_{t,4}$, $\beta(t,s) = \beta_0 + \beta_1 X_{t,4}$, working model $\alpha_0 + \alpha_1 X_{t,1} + \alpha_2 X_{t,4}$, and weight $I_t$. $X_{t,4}$ is included in $S_t$ to assess the effect moderation by $X_{t,4}$, whether the individual received planning support on the previous evening. Table 4 lists the results. There is no evidence of effect moderation by the planning support prompt on the previous day ($\hat{\beta}_1 = 0.046$, $p = 0.734$, 95% CI = -0.228 to 0.320).

## Discussion

In this article we define the causal excursion effect of a digital intervention component in an MRT using the potential outcomes framework. We illustrate how primary and secondary analyses concerning causal excursion effects can be formulated for an MRT, using the HeartSteps MRT as an example. We introduce WCLS as a data analysis method for MRTs, which results in consistent estimators for the causal excursion effect, and describe how to obtain the WCLS estimator through standard statistical software. We illustrate WCLS by using it to analyze the marginal and moderated causal excursion effects using data from the HeartSteps MRT.

**Using Moderation Effect Analysis to Inform JITAI Development**

Conducting moderation analyses, as well as exit interviews with participants, can be useful both in formulating decision rules and in generating hypotheses to be tested in subsequent optimization trials. For example, exit interviews might reveal that participants found that the activity suggestions begin to appear similar as the trial progressed. This combined with the evidence of moderation by day in study might motivate the development of different types of activity suggestions that could be introduced after, say, intervention week 3. The moderating effect of location is an early indication that the decision rules might specify no delivery of activity suggestions when an individual is at the "other" location. In the case of HeartSteps, findings from analyses such as those above, along with other moderation analyses and exit interviews, were used to inform a second MRT currently underway in which a personalization algorithm is being used to reduce the probability of receiving an activity suggestion when there is evidence of a decreasing effect. The conjecture is that intervention effects will stop decaying if the probability of delivering an activity suggestion to an individual is decreased whenever this individual is showing evidence of a decreasing effect. This algorithm is also using location as a moderator.

**Internal and External Validity in MRTs**

Internal validity concerns the ability of the MRT to provide evidence for attributing the estimated effects to the manipulation of the intervention component and not some systematic error (Jüni et al., 2001). It is well known that in a two-arm randomized controlled trial, internal validity is harmed if the randomization, by chance, did not achieve balance in baseline covariates between the two arms. One way to check for deviations that indicate a lack of internal validity is to check whether the distribution of the baseline variables is dissimilar across the two arms. For the MRT, because the randomization occurs sequentially over time, to check internal validity one can check for balance in any covariates occurring prior to each decision point. In the HeartSteps example, one can check whether, for available participants at decision point $t$, the fraction of participants who are at home ($S_t = 1$) is roughly the same among those randomized to an activity suggestion ($A_t = 1$), compared to those randomized to no activity suggestion ($A_t = 0$). Other time-varying variables observed prior to decision point $t$ besides location might be considered as well. Because the causal excursion effect is defined only for individuals who are currently available—that is, one aims to estimate causal effects only among those who are available at decision point $t$—these checks concern only available individuals[7].

External validity concerns the extent to which the estimated causal excursion effect in the MRT provides a basis for generalization to a target population (Jüni et al., 2001). As is well known, in randomized controlled trials external validity is enhanced by striving to enroll participants who are representative of the target population. The same considerations hold in an MRT. One way to assess the extent of external validity (to a defined target population) is to check whether the distribution of the baseline variables is similar or dissimilar to that in the target population. If some baseline variables are likely prognostic for the outcome or predictive for the causal excursion effect, then a distributional imbalance in these variables between the target population and the MRT sample raises concerns that the causal excursion effect estimated from the MRT might not generalize to the target population. If such imbalances are not evident, then greater confidence in the generalizability of the estimated causal excursion effect is justified. In addition to the proximal outcome, $Y_{t+1}$, an MRT can involve other outcomes, such as availability, $I_t$, and the potential moderators, $S_t$. Therefore, any baseline variable that might be related to any of the outcomes should be considered in checking for the aforementioned imbalance.

---

[7] Note that this only applies to MRTs with constant randomization probability (such as the HeartSteps MRT). For MRTs where the randomization probability may change depending on the individual's history information, the aforementioned covariate imbalance may no longer be indicative of lack of internal validity.

The "excursion" aspect of the causal excursion effect is also important when considering generalizability of the findings. The excursion aspect explicitly acknowledges that, prior to decision point $t$, the individual was provided a particular treatment schedule as used in the MRT (rather than some other fixed treatment assignment); the interpretation of the causal excursion effect is the causal effect of excursions from the existing treatment schedule. In the case of the HeartSteps MRT, the existing treatment schedule is "deliver activity suggestion with probability 0.6, if user is available at the decision point" and the excursion effect is a contrast between sending activity message now and not sending activity message now, assuming the user had experienced the existing treatment schedule up to now. The excursion aspect makes it overt that the comparison of two excursions at time $t$ might depend on how treatments were assigned prior to that time, which, in turn, depends on the treatment schedule of the particular MRT. Therefore, the causal excursion effect estimated from an MRT with a particular treatment schedule may differ from the causal excursion effect estimated from an MRT with a different treatment schedule. Recall that the main goal of an MRT is to inform intervention development by identifying ways to improve the *existing* treatment schedule (see the subsection "Using moderation effect analysis to inform intervention development"), and focusing on the causal excursion effect allows the investigator to do exactly that.

**Connection to the Multiphase Optimization Strategy (MOST)**

As discussed in the companion paper (Walton, et al., under review), the MRT fits naturally within the multiphase optimization strategy (MOST; e.g. Collins, 2018; Collins & Kugler, 2018). MOST is an engineering-inspired approach for optimization of behavioral, biobehavioral, and biomedical interventions. In the optimization phase of MOST, the investigator conducts one or more randomized experiments, called optimization trials, aimed at gathering scientific information about the causal effects of individual intervention components, which are needed to construct a new optimized intervention or to optimize an existing intervention. Typically, this includes inference for causal effects of individual intervention components, as well as moderation analyses. The MRT is one design that can be used for an optimization trial, along with factorial and fractional factorial designs (Collins et al., 2009), the sequential, multiple assignment, randomized trial (SMART; Collins, Nahum-Shani, & Almirall, 2014), and other experimental designs. This paper defines inferential methods for the causal effects in an MRT in the optimization phase of MOST.

Inference concerning causal excursion effects fits naturally within the overall conceptual framework of MOST. In this framework optimization is an ongoing process of intervention improvement, in which each optimization trial provides information useful in generating hypotheses about how to improve the intervention further and, therefore, informs the design of the next optimization trial. For example, the following question can be characterized by the causal excursion effect: If the treatment schedule for the activity suggestions based on knowledge of the current location were altered, would this improve subsequent 30-minute step count? In digital interventions this inferential goal makes sense even in implementation as the team must continually monitor and update the digital application software. Similarly, continually monitoring performance and assessing how to best improve the current schedule for assigning treatments is natural. The causal excursion effect is useful for this purpose.

**Sample Size Considerations**

Liao et al. (2016) provides theory for determining the sample size for the setting in which the moderator $S_t$ is exogenous (for example, time since the individual started the intervention). A web applet can be found at https://methodologycenter.shinyapps.io/mrt_ss/. The applet takes as input duration of the study, number of decision points per day, expected availability pattern, randomization probability (which can be time-varying), target proximal treatment effect to be detected, and the desired power, and it outputs the required sample size (or vice versa; input sample size and output power). Alternately, R code is freely available at https://cran.r-project.org/web/packages/MRTSampleSize/index.html.

**Additional Types of Causal Effects**

This paper focuses on the immediate causal excursion effect ("immediate" in the sense that there is no other treatment between the decision point $t$ and the proximal outcome $Y_{t+1}$) of a time-varying digital intervention. One may also be interested in inference about a delayed causal excursion effect. For example, when assessing the effect of the planning support component, it may be of interest to assess the effect of a planning support prompt on the total step count over the next $x$ days, as it would be desirable for the delivery of a planning prompt to have a longer-term effect, such as forming a habit. The generalization of WCLS to assess such delayed effects is given in Boruvka et al. (2018). Here we note only that the interpretation of such delayed causal excursion effects averages over, in addition to the history information observed up to that decision point, future treatments and future covariates. Suppose researchers are interested in the effect of a planning support prompt on the total step count over the next $x$ days. This effect would be marginal with respect to the schedule for delivering planning support during the subsequent $x$ days.

Other more familiar causal effects might be estimated, but additional assumptions are necessary. For example, suppose it can be safely assumed that the treatments prior to the current decision point will not impact subsequent outcomes (i.e., these prior treatments do not have delayed positive or negative effects). Then the potential outcomes such as $\left(Y_{t+1}(\bar{a}_t), I_t(\bar{a}_{t-1}), S_t(\bar{a}_{t-1})\right)$ are actually $\left(Y_{t+1}(a_t), I_t(a_{t-1}), S_t(a_{t-1})\right)$, making it reasonable to focus on inference for the effect, $E[Y_{t+1}(1) - Y_{t+1}(0) | I_t(A_{t-1}) = 1, S_t(A_{t-1}) = s]$. In terms of the primary analysis of data from an MRT, we opt to make inference about causal excursion effects due to both its interpretation in the above continual learning paradigm and the minimal causal inference assumptions it requires. Of course, in secondary and hypothesis-generating analyses, a variety of statistical assumptions would be made to draw inferences about other causal effects.

**Other Types of Analyses**

Generalized estimating equations (GEE; Liang & Zeger, 1986) and multi-level models (MLM; Laird & Ware, 1982; Raudenbush & Bryk, 2002) have been used with great success to analyze data from intensive longitudinal studies; at first glance they appear to be a natural choice for conducting primary and secondary data analysis for MRTs. However, these methods can result in biased causal effect estimates for MRTs when there are endogenous time-varying covariates—covariates that can depend on previous outcomes or previous treatments. For example, in HeartSteps, the prior 30-minute step count is likely impacted by prior treatment and is thus endogenous. We illustrate this bias in Appendix A.

**Limitations and Future Directions**

Limitations of the WCLS method presented in this paper include the following. (i) The method presented here is applicable only to the case where the proximal outcome is continuous. Qian, Yoo, Klasnja, Almirall, & Murphy (2019) provide an alternative method for analyzing MRT data with a binary outcome that yields causal effect estimates on the relative risk scale. (ii) The method presented here provides estimates of marginal effects; thus it does not provide person-specific predictions of treatment effect except as explained by observed covariates.

Future directions include the development of (i) related methods for other types of proximal outcomes (e.g., zero-inflated, categorical, ordinal, and longitudinal proximal outcomes), and (ii) multi-level models to model person-specific causal excursion effects that are appropriate for MRTs.

## Appendix A

## GEE and MLM Can Be Biased When Estimating Causal Excursion Effects in MRTs

MRTs produce intensive longitudinal data (Schafer, 2006), as individuals are randomized among intervention options repeatedly during the MRT, and outcomes and covariates are assessed in tandem with randomization. Repeated measurement of the same individuals over time means that the repeated observations are likely dependent. *Generalized estimating equations* (GEE; Liang & Zeger, 1986) and *multi-level models* (MLM; Laird & Ware, 1982; Raudenbush & Bryk, 2002), the latter also known as mixed models or random effects models, have been used widely in analyzing longitudinal data. However, as we illustrate below, inappropriate application of them to MRT data may result in biased estimates of the causal excursion effects when *endogenous time-varying covariates* are included in the model. A time-varying covariate is *endogenous* if it can depend on previous outcomes or previous treatments, which commonly occurs in MRTs. For example, in analyzing the effect of activity suggestion in the subsequent 30-minute step count in HeartSteps, one may want to control for the 30-minute step count prior to each decision point to reduce noise. Because the 30-minute step count prior to a decision point can be correlated with past step counts (i.e., past outcomes), it is endogenous. When a time-varying covariate is not endogenous, it is called *exogenous*. Examples of *exogenous time-varying covariates* include time, weather, and anything that cannot be impacted by previous treatments or previous outcomes.

**Inappropriate Use of GEE and MLM Can Result in Biased Causal Excursion Effect Estimates in the Presence of Endogenous Time-Varying Covariates**

Pepe & Anderson (1994) demonstrated that, in the presence of endogenous time-varying covariates, parameter estimates from GEE may be biased unless certain conditions, described below, are met. Such bias is also shown in subsequent research through simulation studies and analytic calculations (Diggle et al., 2002; Pan et al., 2000; Schildcrout & Heagerty, 2005; Tchetgen et al., 2012; Vansteelandt, 2007). For completeness we provide a brief explanation of the bias here. Consider a simplified version of the HeartSteps MRT, where there are two decision points for each individual and individuals are always available. Suppose the observed data for individual $i$ is $(X_{i1}, A_{i1}, Y_{i2}, X_{i2}, A_{i2}, Y_{i3})$, where $X_{it}$ denotes the 30-minute step count prior to decision point $t$ (an endogenous time-varying covariate), $A_{it}$ is the indicator of whether an activity suggestion is delivered at decision point $t$ (so $A_{it}$ has .6 probability to be 1), and $Y_{i,t+1}$ is the 30-minute step count following decision point $t$. The researcher chooses $S_{it} = X_{it}$ in equation (2): they want to assess whether the effect of the activity suggestion is moderated by the prior 30-minute step count. The researcher may then choose to impose the following linear model on the mean of the proximal outcome given the treatment and the covariate at decision point $t$:

$$E(Y_{i,t+1} | A_{it}, X_{it}) = \alpha_0 + \alpha_1 X_{it} + A_{it}(\beta_0 + \beta_1 X_{it}), \tag{7}$$

and use GEE to estimate the coefficients $\alpha_0, \alpha_1, \beta_0, \beta_1$.[8] Often a non-independent working correlation structure is used in GEE, aiming for efficiency gain (i.e., smaller standard error of the estimated coefficients compared to GEE with working independence correlation structure).

It is well known that GEE produces consistent estimates regardless of the choice of the working correlation structure, as long as equation (7) holds; however, this is only true when all covariates are exogenous. In this above example with two decision points, Pepe & Anderson (1994) demonstrated that to guarantee the consistency of the GEE estimates, one of the following conditions needs to hold:

(i) $E(Y_{i,t+1}|A_{it}, X_{it}) = E(Y_{i,t+1}|A_{i1}, X_{i1}, A_{i2}, X_{i2})$ for $t = 1,2$; or

(ii) a working independence correlation structure is used.

---

[8] In this particular example in which the randomization probability is constant, the $\beta_0, \beta_1$ in the proximal treatment effect term in (7) equals the $\beta_0', \beta_1'$ in $E[E(Y_{t+1} | A_t = 1, H_t) - E(Y_{t+1} | A_t = 0, H_t) | S_t] = \beta_0' + \beta_1' S_t$. Therefore, if one can obtain consistent estimates for $\beta_0, \beta_1$, one obtains consistent estimate for the proximal treatment effect defined in (2). In general, however, when the randomization probability can depend on $H_t$, the $\beta_0, \beta_1$ in (7) no longer equals the $\beta$ in (2) due to the marginalization over $S_t$. This is another reason, in addition to the reason that will be presented in the next paragraph of the paper, why inappropriate use of GEE results in biased proximal treatment effect estimates.

Condition (i) is usually violated when $X_{it}$ is endogenous: In this particular example, $X_{i2}$ can be correlated with $Y_{i2}$, so that $E(Y_{i2}|A_{i1}, X_{i1}) \neq E(Y_{i2}|A_{i1}, X_{i1}, A_{i2}, X_{i2})$. This means that unless the independent working correlation structure is used, GEE can produce biased estimates even if equation (7) holds.

The same bias can occur when MLM is used instead of GEE. In general, for each MLM there is a corresponding GEE with a non-independent correlation structure that produces the same estimated coefficients. For example, an MLM resembling equation (7) is $Y_{i,t+1} = \alpha_0 + \alpha_1 X_{it} + A_{it}(\beta_0 + \beta_1 X_{it}) + u_i + \epsilon_{it}$, where $u_i \sim \text{Normal}(0, \sigma_u^2)$ is a random intercept and $\epsilon_{it} \sim \text{Normal}(0, \sigma_\epsilon^2)$ is the error term. This corresponds to a GEE with compound symmetric (also called exchangeable) working correlation structure. Given this equivalency, MLM can produce biased estimates if the covariate $X_{it}$ is endogenous.

**A Few Scenarios Where GEE or MLM Provides Consistent Causal Excursion Effect Estimates From MRT Data**

GEE builds upon a marginal mean model (i.e., the relationship between the mean of the proximal outcome, the covariates, and the treatment assignments, such as (7)). If no endogenous time-varying covariates are included in the model, the individuals are always available, and the randomization probability is constant, GEE with any working correlation structure gives consistent estimates as long as the marginal mean model is correct. If there are endogenous time-varying covariates in the model, the individuals are always available, and the randomization probability is constant, GEE with independent working correlation structure still gives consistent estimates as long as the marginal mean model is correct, but GEE with other working correlation structure does not.

Because an MLM always corresponds to a GEE with some non-independent working correlation structure, MLM provides consistent causal excursion effect estimates if no endogenous time-varying covariates are included in the model, the individuals are always available, and the randomization probability is constant. However, although the estimated coefficients from an MLM will generally be biased for the causal excursion effect when there are endogenous time-varying covariates, those estimated coefficients can have a different, individual-specific interpretation under a rather strong assumption. As shown in Qian, Klasnja, & Murphy (2019), if the endogenous time-varying covariates can be safely assumed to only depend on the random effect through the observed previous outcomes and previous covariates, then the fitted results from standard linear mixed models can be interpreted as a causal effect that is conditional on the random effect (i.e., individual-specific rather than population-average) and conditional on the entire history $H_t$ (rather than conditional only on $S_t$). An example where this strong assumption holds is when the endogenous time-varying covariates are previous proximal outcomes (e.g., the endogenous time-varying covariate at decision point $t$ is the proximal outcome at decision point $t-1$).

## A Mathematical Demonstration of the Bias From Inappropriate Application of GEE When There are Endogenous Time-Varying Covariates

For clarity we consider the case where each participant is in the MRT for two decision points. The data for the $i$-th participant is $(X_{i1}, A_{i1}, Y_{i2}, X_{i2}, A_{i2}, Y_{i3})$, where $X_{it}$ is the covariate, $A_{it}$ is the treatment assignment, and $Y_{it+1}$ is the continuous outcome. The covariate $X_{it}$ is endogenous time-varying, in the sense that it can depend on previous treatment and previous outcome.

The model on the marginal mean of $Y_{t+1}$ is $E(Y_{t+1}|A_t, X_t) = \alpha_0 + \alpha_1 X_t + A_t(\beta_0 + \beta_1 X_t)$. The corresponding GEE solves the following estimating equation:

$$\sum_{i=1}^{n} \begin{bmatrix} 1 & 1 \\ X_{i1} & X_{i2} \\ A_{i1} & A_{i2} \\ A_{i1}X_{i1} & A_{i2}X_{i2} \end{bmatrix} V^{-1} \begin{bmatrix} Y_{i2} - \alpha_0 - \alpha_1 X_{i1} - A_{i1}(\beta_0 + \beta_1 X_{i1}) \\ Y_{i3} - \alpha_0 - \alpha_1 X_{i2} - A_{i2}(\beta_0 + \beta_1 X_{i2}) \end{bmatrix} = 0. \quad (8)$$

Here, $n$ denotes the number of participants, and $V$ is a $2 \times 2$ working covariance matrix. Examples of $V$ include the following:

- Working independence: $V = \begin{bmatrix} \sigma^2 & 0 \\ 0 & \sigma^2 \end{bmatrix}$
- Compound symmetry: $V = \begin{bmatrix} \sigma^2 & \rho\sigma^2 \\ \rho\sigma^2 & \sigma^2 \end{bmatrix}$
- Autoregressive (in the special case of two decision points, autoregressive is the same as compound symmetry): $V = \begin{bmatrix} \sigma^2 & \rho\sigma^2 \\ \rho\sigma^2 & \sigma^2 \end{bmatrix}$.

In this setting, the result in Pepe & Anderson (1994) implies that GEE is guaranteed to produce consistent $\alpha_0, \alpha_1, \beta_0, \beta_1$ if either

(i) $E(Y_{t+1}|A_t, X_t) = E(Y_{t+1}|A_1, X_1, A_2, X_2)$ for $t = 1,2$, or

(ii) a working independence correlation structure is used,

and they provided simulation results to show that GEE can produce biased estimates when neither condition holds. In the following, we rephrase the intuitive argument given in Pepe and Anderson (1994) in this particular setting to show why GEE can be biased if neither condition holds.

We write $V^{-1} = \begin{bmatrix} w_{11} & w_{12} \\ w_{21} & w_{22} \end{bmatrix}$ and write the residual $r_{it} = Y_{it+1} - \alpha_0 - \alpha_1 X_{it} - A_{it}(\beta_0 + \beta_1 X_{it})$. A summand (for fixed $i$) in equation (8) becomes

$$\begin{bmatrix} 1 & 1 \\ X_{i1} & X_{i2} \\ A_{i1} & A_{i2} \\ A_{i1}X_{i1} & A_{i2}X_{i2} \end{bmatrix} \begin{bmatrix} w_{11} & w_{12} \\ w_{21} & w_{22} \end{bmatrix} \begin{bmatrix} r_{i1} \\ r_{i2} \end{bmatrix}$$
$$= \begin{bmatrix} (w_{11} + w_{21})r_{i1} + (w_{12} + w_{22})r_{i2} \\ (w_{11}X_{i1} + w_{21}X_{i2})r_{i1} + (w_{12}X_{i1} + w_{22}X_{i2})r_{i2} \\ (w_{11}A_{i1} + w_{21}A_{i2})r_{i1} + (w_{12}A_{i1} + w_{22}A_{i2})r_{i2} \\ (w_{11}A_{i1}X_{i1} + w_{21}A_{i2}X_{i2})r_{i1} + (w_{12}A_{i1}X_{i1} + w_{22}A_{i2}X_{i2})r_{i2} \end{bmatrix}. \quad (9)$$

Because $E(Y_{t+1}|A_t, X_t) = \alpha_0 + \alpha_1 X_t + A_t(\beta_0 + \beta_1 X_t)$, we have

$$E[r_{it}] = E[X_{it}r_{it}] = E[A_{it}r_{it}] = E[A_{it}X_{it}r_{it}] = 0.$$

Therefore, all the terms with $w_{11}r_{i1}$ and $w_{22}r_{i2}$ (such as $w_{11}r_{i1}X_{i1}$; i.e., terms that are multiplied with the diagonal elements of $V^{-1}$) in (9) have expectation zero, and what is left are the terms with $w_{21}r_{i1}$ and $w_{12}r_{i2}$ (i.e., terms that are multiplied with the off-diagonal elements of $V^{-1}$). In other words, the expectation of (9) equals

$$\begin{bmatrix} 0 \\ w_{21}X_{i2}r_{i1} + w_{12}X_{i1}r_{i2} \\ w_{21}A_{i2}r_{i1} + w_{12}A_{i1}r_{i2} \\ w_{11}A_{i2}X_{i2}r_{i1} + w_{12}A_{i1}X_{i1}r_{i2} \end{bmatrix}. \quad (10)$$

Mathematical theory for GEE tells us that GEE outputs consistent $\alpha_0, \alpha_1, \beta_0, \beta_1$ when (9) has expectation zero; i.e., when (10) equals zero.

If condition (i) holds, we have $E[X_{i2}r_{i1}] = E[A_{i2}r_{i1}] = E[A_{i2}X_{i2}r_{i1}] = 0$, and similarly $E[X_{i1}r_{i2}] = E[A_{i1}r_{i2}] = E[A_{i1}X_{i1}r_{i2}] = 0$. Therefore, (10) equals 0 with any choice of $V^{-1}$, and GEE estimators are consistent.

If condition (ii) holds, we have $w_{21} = w_{12} = 0$. Hence (10) equals 0 and GEE estimators are consistent.

When neither condition holds, it's likely that (10) does not equal zero. For example, suppose $X_{i2} = Y_{i2}$. Then the term $X_{i2}r_{i1}$ equals

$$Y_{i2}\{Y_{i2} - \alpha_0 + \alpha_1 X_{i1} + A_{i1}(\beta_0 + \beta_1 X_{i1})\}, \quad (11)$$

which is the residual multiplied with the outcome itself. Because the residual and the outcome at the same time point are correlated, (11) likely does not equal zero. Therefore, (10) likely does not equal zero. This means GEE can be biased when neither conditions hold, i.e., when endogenous time-varying covariates are included and non-independent working correlation structure is used.

## Appendix B

## A General Form of the WCLS Estimator for the Causal Excursion Effect That Allows the Randomization Probability to Vary Over Time

We assume a linear model for the causal excursion effect: $\beta(t, s) = s^T \beta$. Suppose $Z_t^T \alpha$ is a working model for the conditional mean of $Y_{t+1}$ given no treatment at decision point t and history $H_t$, $E(Y_{t+1}|A_t = 0, I_t = 1, H_t)$. Note that the consistency of the estimator for $\beta$ does not require $Z_t^T \alpha$ to be a correct model for $E(Y_{t+1}|A_t = 0, I_t = 1, H_t)$. We use $p_t(H_t)$ to denote the randomization probability at decision point $t$, which may possibly depend on $H_t$.

The WCLS estimator for $\beta$ is calculated as follows. Suppose $(\hat{\alpha}, \hat{\beta})$ is the $(\alpha, \beta)$ value that solves the following estimating equation:

$$\sum_{i=1}^{n} \sum_{t=1}^{T} I_{it} W_{it} \left[ Y_{i,t+1} - Z_{it}^T \alpha - \{A_{it} - \tilde{p}_t(S_{it})\} S_{it}^T \beta \right] \begin{bmatrix} Z_{it} \\ \{A_{it} - \tilde{p}_t(S_{it})\} S_{it} \end{bmatrix} = 0; \quad (12)$$

then $\hat{\beta}$ is the WCLS estimator for $\beta$. $\tilde{p}_t(S_{it})$ is an arbitrary probability as long as it depends on $H_{it}$ through at most $S_{it}$ and it is bounded away from 0 and 1; $i$ is the index for the $i$th individual, and $W_{it}$ is defined as

$$W_{it} = \left\{ \frac{\tilde{p}_t(S_{it})}{p_t(H_{it})} \right\}^{A_{it}} \left\{ \frac{1 - \tilde{p}_t(S_{it})}{1 - p_t(H_{it})} \right\}^{1 - A_{it}}. \quad (13)$$

$W_{it}$, the ratio of two probabilities, serves as a change of probability: It makes it as if the treatment $A_{it}$ is randomized with probability $\tilde{p}_t(S_{it})$. It is used to marginalize the causal excursion effect over variables in $H_{it}$ but not in $S_{it}$. As long as $\tilde{p}_t(S_t)$ depends on $H_{it}$ through at most $S_{it}$ and it is bounded away from 0 and 1, the particular choice of $\tilde{p}_t(S_t)$ doesn't affect the consistency of $\hat{\beta}$. For instance, one can set it to be 0.5 (or any constant between 0 and 1) for all individuals and all decision points, or set it to be the predicted value from a logistic regression fit of $A_t \sim S_t$. If the true randomization probability $p_t(H_t)$ depends at most on $S_t$, then one can also set $\tilde{p}_t(S_t)$ to be equal to the true randomization probability, in which case (12) is mathematically equivalent to (5). $\tilde{p}_t(S_t)$ can impact the standard error of $\hat{\beta}$. In addition, when the causal excursion effect model $\beta(t,s) = s^T\beta$ is misspecified, $\tilde{p}_t(S_t)$ impacts the limit of $\hat{\beta}$. See the Appendix of Boruvka et al. (2018) for more technical details on how the limit of $\hat{\beta}$ is impacted by $\tilde{p}_t(S_t)$ in this case.

Now we present a way to obtain the general WCLS estimator for time-varying randomization probability through standard statistical software that implements GEE. Suppose the assumed causal excursion effect model is (6) and the working model for $E(Y_{t+1}|A_t = 0, I_t = 1, H_t)$ is $Z_t^T\alpha$; then the WCLS estimator $\hat{\beta}$ and its standard error can be obtained by (i) incorporating $I_t W_t$ as the "prior weights", (ii) chooseing a working independence correlation structure, and (iii) fitting GEE with dependent variable $Y_{t+1}$ and independent variables $Z_t$ and $(A_t - \tilde{p}_t(S_t))S_t$. Then the estimated coefficient for $(A_t - \tilde{p}_t(S_t))S_t$ is the WCLS estimate $\hat{\beta}$.

**Standard Error Formula for WCLS.**

Below we provide the formula for the standard error of the WCLS estimator $\hat{\beta}$. For $(\hat{\alpha}, \hat{\beta})$ that solves estimating equation (12), variance can be estimated by

$$\widehat{\text{Var}}\left(\begin{bmatrix}\hat{\alpha}\\\hat{\beta}\end{bmatrix}\right) = \frac{1}{n} M_n^{-1} \Sigma_n (M_n^{-1})^T,$$

where

$$M_n = -\mathbb{P}_n \sum_{t=1}^{T} W_t \begin{bmatrix} Z_t Z_t^T & \{A_t - \tilde{p}_t(S_t)\} Z_t S_t^T \\ \{A_t - \tilde{p}_t(S_t)\} S_t Z_t^T & \{A_t - \tilde{p}_t(S_t)\}^2 S_t S_t^T \end{bmatrix}$$

and

$$\Sigma_n = \mathbb{P}_n \sum_{t=1}^{T} \{Y_{t+1} - Z_t^T \alpha - (A_t - \tilde{p}_t(S_t)) S_t^T \beta\}^2 W_t \begin{bmatrix} Z_t Z_t^T & \{A_t - \tilde{p}_t(S_t)\} Z_t S_t^T \\ \{A_t - \tilde{p}_t(S_t)\} S_t Z_t^T & \{A_t - \tilde{p}_t(S_t)\}^2 S_t S_t^T \end{bmatrix}.$$

Here, $\mathbb{P}_n$ denotes sample average over *n* individuals. The standard error formula can be modified for the setting in which the randomization probability is constant over time (i.e., the setting in the main paper) by letting $\tilde{p}_t(S_t) = p$ and $W_t = 1$.

## Appendix C

We conduct a simulation study to illustrate the claim that including variables that are correlated with $Y_{t+1}$ in $Z_t$ may reduce the variance of the WCLS estimator. The generative model mimics features of the HeartSteps data and is set up as follows. For simplicity we assume users are always available. At decision point $t$, the covariate $X_t$ is drawn from the empirical distribution of the log-transformed 30-minute step count preceding a decision point in the HeartSteps data. For simplicity $X_t$ is generated independently of previous outcomes and treatments. The treatment $A_t$ is generated from a Bernoulli distribution with .6 success probability; this mimics the .6 randomization probability of activity suggestions in HeartSteps. The proximal outcome $Y_{t+1}$ is generated from a Gaussian distribution with mean

$$1.6085 + 0.4037 \times X_t + 0.0655 \times Y_t + 0.1229 \times (A_t - 0.6)$$

and standard deviation 2.716. The coefficients in the above display are the estimated coefficients from a WCLS fit on the HeartSteps data with the same control variables $(1, X_t, Y_t)$ and constant treatment effect model. The standard deviation is the empirical standard deviation of the residual in $Y_{t+1}$ from the above WCLS fit. As in the HeartSteps data set, for each simulated trial we generate 37 individuals, each with 210 decision points.

For each data set generated from the above generative model, we consider four WCLS fits for the true treatment effect 0.1229 and compare their performance. All four WCLS assume the constant treatment effect model, and they differ in the choice of the working model. The first WCLS fit (WCLS-1) includes control variables $(1, X_t, Y_t)$; the second WCLS fit (WCLS-2) includes control variables $(1, X_t)$; the third WCLS fit (WCLS-3) includes control variables $(1, Y_t)$; and the fourth WCLS fit (WCLS-4) includes only the intercept. The bias, standard deviation (SD), and coverage probability (CP) of 95% confidence interval are listed in Supplementary Table 1. All four WCLS estimators are consistent with nominal confidence interval coverage because their assumed constant treatment effect model holds under this generative model. (This again illustrates that the consistency of the WCLS estimator does not require the control part of the model to be correct.) On the other hand, the choice of working model affects the efficiency of the estimator. In particular, WCLS-1 and WCLS-2 have smaller standard errors than WCLS-3 and WCLS-4 because the former two include $X_t$, a covariate that is highly correlated with the proximal outcome $Y_{t+1}$.

**Appendix D**

To assess sensitivity of the result to potential non-linearity in Question 2 of Section "Analysis Using Data from HeartSteps MRT," we fit a local 2-degree polynomial regression with smoothing span 2/3 and tricubic weighting to estimate the causal excursion effect over time (the default setting for many local regression software, such as the lowess function in R (R Core Team, 2019)). The estimated effect from local regression is presented in Supplementary Figure 1 (black curve). Comparing this estimated effect with the estimated effect based on the linear model (blue curve in Figure 1, with blue shaded area being the pointwise 95% confidence interval), we see that the two estimates are relatively close to each other, indicating that the linear model fits well.

**References**

Boruvka, A., Almirall, D., Witkiewitz, K., & Murphy, S. A. (2018). Assessing time-varying causal effect moderation in mobile health. *Journal of the American Statistical Association*, *113*(523), 1112–1121.

Clawson, J., Pater, J. A., Miller, A. D., Mynatt, E. D., & Mamykina, L. (2015). No longer wearing: Investigating the abandonment of personal health-tracking technologies on craigslist. *UbiComp 2015 - Proceedings of the 2015 ACM International Joint Conference on Pervasive and Ubiquitous Computing*, 647–658. https://doi.org/10.1145/2750858.2807554

Collins, L. M., Dziak, J. J., & Li, R. (2009). Design of experiments with multiple independent variables: A resource management perspective on complete and reduced factorial designs. *Psychological Methods*, *14*(3), 202–224. https://doi.org/10.1037/a0015826

Dempsey, W., Liao, P., Kumar, S., & Murphy, S. A. (2019). The stratified micro-randomized trial design: Sample size considerations for testing nested causal effects of time-varying treatments. *ArXiv: 1711.03587*.

Diggle, P., Heagerty, P., Liang, K.-Y., & Zeger, S. (2002). *Analysis of longitudinal data*. Oxford University Press.

Eysenbach, G. (2005). The law of attrition. *Journal of Medical Internet Research*, *7*(1), 1–11. https://doi.org/10.2196/jmir.7.1.e11

Ho, J., & Intille, S. S. (2005). Using context-aware computing to reduce the perceived burden of interruptions from mobile devices. *CHI 2005: Technology, Safety, Community: Conference Proceedings - Conference on Human Factors in Computing Systems*, 909–918.

Holland, P. W. (1986). Statistics and causal inference. *Journal of the American Statistical Association*, *81*(396), 945–960.

Hong, G., & Raudenbush, S. W. (2006). Evaluating kindergarten retention policy: A case study of causal inference for multilevel observational data. *Journal of the American Statistical Association*, *101*(475), 901–910. https://doi.org/10.1198/016214506000000447

Hudgens, M. G., & Halloran, M. E. (2008). Toward causal inference with interference. *Journal of the American Statistical Association*, *103*(482), 832–842. https://doi.org/10.1198/016214508000000292

IBM Corp. (2019). *IBM SPSS Statistics for Windows*.

Jüni, P., Altman, D., & Egger, M. (2001). Assessing the quality of controlled clinical trials. *BMJ*, *323*(7303), 42–46.

Klasnja, P., Harrison, B. L., Legrand, L., Lamarca, A., Froehlich, J., & Hudson, S. E. (2008). Using wearable sensors and real time inference to understand human recall of routine activities. *UbiComp 2008 - Proceedings of the 10th International Conference on Ubiquitous Computing*, 154–163. https://doi.org/10.1145/1409635.1409656

Klasnja, P., Hekler, E. B., Shiffman, S., Boruvka, A., Almirall, D., Tewari, A., & Murphy, S. A.

(2015). Microrandomized trials: An experimental design for developing just-in-time adaptive interventions. *Health Psychology*, *34*(S), 1220.

Klasnja, P., Smith, S., Seewald, N. J., Lee, A., Hall, K., Luers, B., Hekler, E. B., & Murphy, S. A. (2018). Efficacy of contextually tailored suggestions for physical activity: A micro-randomized optimization trial of HeartSteps. *Annals of Behavioral Medicine : A Publication of the Society of Behavioral Medicine*, *53*(6), 573–582. https://doi.org/10.1093/abm/kay067

Laird, N. M., & Ware, J. H. (1982). Random-effects models for longitudinal data. *Biometrics*, *38*(4), 963–974.

Liang, K.-Y., & Zeger, S. L. (1986). Longitudinal data analysis using generalized linear models. *Biometrika*, *73*(1), 13–22.

Liao, P., Klasnja, P., Tewari, A., & Murphy, S. A. (2016). Sample size calculations for micro-randomized trials in mHealth. *Statistics in Medicine*, *35*(12), 1944–1971.

Mancl, L. A., & DeRouen, T. A. (2001). A covariance estimator for GEE with improved small-sample properties. *Biometrics*, *57*(1), 126–134. https://doi.org/10.1111/j.0006-341X.2001.00126.x

Nahum-Shani, I., Smith, S. N., Spring, B. J., Collins, L. M., Witkiewitz, K., Tewari, A., & Murphy, S. A. (2018). Just-in-time adaptive interventions (JITAIs) in mobile health: Key components and design principles for ongoing health behavior support. *Annals of Behavioral Medicine*, *52*(6), 446–462.

Pan, W., Louis, T. A., & Connett, J. E. (2000). A note on marginal linear regression with correlated response data. *American Statistician*, *54*(3), 191–195. https://doi.org/10.1080/00031305.2000.10474544

Pepe, M. S., & Anderson, G. L. (1994). A cautionary note on inference for marginal regression models with longitudinal data and general correlated response data. *Communications in Statistics-Simulation and Computation*, *23*(4), 939–951.

Qian, T., Klasnja, P., & Murphy, S. A. (2019). Linear mixed models under endogeneity: Modeling sequential treatment effects with application to a mobile health study. *Statistical Science*.

Qian, T., Yoo, H., Klasnja, P., Almirall, D., & Murphy, S. A. (2019). Estimating time-varying causal excursion effect in mobile health with binary outcomes. *ArXiv: 1906.00528*.

R Core Team. (2019). *R: A language and environment for statistical computing*. https://www.r-project.org

Rankin, C. H., Abrams, T., Barry, R. J., Bhatnagar, S., Clayton, D. F., Colombo, J., Coppola, G., Geyer, M. A., Glanzman, D. L., Marsland, S., McSweeney, F. K., Wilson, D. A., Wu, C. F., & Thompson, R. F. (2009). Habituation revisited: An updated and revised description of the behavioral characteristics of habituation. *Neurobiology of Learning and Memory*, *92*(2), 135–138. https://doi.org/10.1016/j.nlm.2008.09.012

Raudenbush, S. W., & Bryk, A. S. (2002). *Hierarchical linear models: Applications and data analysis methods* (Vol. 1). Sage.

Robins, J. M. (1986). A new approach to causal inference in mortality studies with a sustained exposure period—application to control of the healthy worker survivor effect. *Mathematical Modelling*, *7*(9–12), 1393–1512.

Robins, J. M. (1987). Addendum to "a new approach to causal inference in mortality studies with a sustained exposure period—application to control of the healthy worker survivor effect." *Computers & Mathematics with Applications*, *14*(9–12), 923–945.

Robins, J. M. (1994). Correcting for non-compliance in randomized trials using structural nested mean models. *Communications in Statistics - Theory and Methods*, *23*(8), 2417–2421. https://doi.org/10.1080/03610929408831394

Robins, J. M., Hernán, M. Á., & Brumback, B. (2000). Marginal structural models and causal inference in epidemiology. *Epidemiology*, *11*(5), 550–560. https://doi.org/10.1097/00001648-200009000-00011

Rubin, D. B. (1978). Bayesian inference for causal effects: The role of randomization. *The Annals of Statistics*, 34–58.

Rubin, D. B. (2005). Causal inference using potential outcomes: Design, modeling, decisions. *Journal of the American Statistical Association*, *100*(469), 322–331.

SAS Institute Inc. (2019). *SAS/STAT Software, Version 9.4*. http://www.sas.com/

Schildcrout, J. S., & Heagerty, P. J. (2005). Regression analysis of longitudinal binary data with time-dependent environmental covariates: Bias and efficiency. *Biostatistics*, *6*(4), 633–652. https://doi.org/10.1093/biostatistics/kxi033

Shaw, R. J., Bosworth, H. B., Hess, J. C., Silva, S. G., Lipkus, I. M., Davis, L. L., & Johnson, C. M. (2013). Development of a theoretically driven mHealth text messaging application for sustaining recent weight loss. *Journal of Medical Internet Research*, *15*(5). https://doi.org/10.2196/mhealth.2343

StataCorp. (2019). *Stata Statistical Software: Release 16*.

Tchetgen, E. J. T., Glymour, M. M., Weuve, J., & Robins, J. (2012). Specifying the correlation structure in inverse-probability-weighting estimation for repeated measures. *Epidemiology*, *23*(4), 644–646. https://doi.org/10.1097/EDE.0b013e31825727b5

Vansteelandt, S. (2007). On confounding, prediction and efficiency in the analysis of longitudinal and cross-sectional clustered data. *Scandinavian Journal of Statistics*, *34*(3), 478–498. https://doi.org/10.1111/j.1467-9469.2006.00555.x

Walls, T. A., & Schafer, J. L. (2006). *Models for intensive longitudinal data*. Oxford University Press.

Walton, A., Collins, L. M., Corsby, L., Klasnja, P., Nahum-Shani, I., Rabbi, M., Walton, M., & Murphy, S. A. The micro-randomized trial for developing mobile health interventions. *Under Review*.

Yardley, L., Spring, B. J., Riper, H., Morrison, L. G., Crane, D. H., Curtis, K., Merchant, G. C., Naughton, F., & Blandford, A. (2016). Understanding and promoting effective engagement with digital behavior change interventions. *American Journal of Preventive Medicine*, *51*(5), 833–842. https://doi.org/10.1016/j.amepre.2016.06.015

Table 1.

*Estimated main effect of activity suggestions on proximal outcome*

| Variable | | Estimate | 95% LCL | 95% UCL | SE | Hotelling $t$ | $p$ |
|---|---|---|---|---|---|---|---|
| Intercept | $\alpha_0$ | 1.783 | 1.537 | 2.029 | 0.121 | 217.3 | <0.001 |
| Past 30-min step count | $\alpha_1$ | 0.414 | 0.351 | 0.476 | 0.031 | 181.2 | <0.001 |
| Activity suggestion | $\beta_0$ | 0.131 | -0.006 | 0.268 | 0.067 | 3.79 | 0.060 |

*Note.* LCL (UCL) represents lower (upper) confidence limit. SE represents standard error. LCL, UCL, SE, and *p* are corrected for small sample size using method in (Liao et al., 2016; Mancl & DeRouen, 2001). The degrees of freedom for the Hotelling *t* test is (1, 34).

Table 2.

*Estimated effect of activity suggestion on proximal outcome as a linear function of time in study*

| Variable | | Estimate | 95% LCL | 95% UCL | SE | Hotelling $t$ | $p$ |
|---|---|---|---|---|---|---|---|
| Intercept | $\alpha_0$ | 2.003 | 1.765 | 2.240 | 0.117 | 294.7 | <0.001 |
| Past 30-minute step count | $\alpha_1$ | 0.412 | 0.351 | 0.473 | 0.030 | 189.6 | <0.001 |
| Time (in days) | $\alpha_2$ | -0.011 | -0.020 | -0.001 | 0.005 | 5.09 | 0.031 |
| Activity suggestion | $\beta_0$ | 0.507 | 0.201 | 0.814 | 0.151 | 11.37 | 0.002 |
| Activity suggestion x Time (in days) | $\beta_1$ | -0.018 | -0.031 | -0.006 | 0.006 | 9.19 | 0.005 |

*Note.* LCL (UCL) represents lower (upper) confidence limit. SE represents standard error. LCL, UCL, SE, and *p* are corrected for small sample size using method in (Liao et al., 2016; Mancl & DeRouen, 2001). The degrees of freedom for the Hotelling *t* test is (1, 32).

Table 3.

*Estimated effect of walking suggestion / anti-sedentary suggestion on proximal outcome, moderated by location (home/work or other) during decision point*

| Variable | | Estimate | 95% LCL | 95% UCL | SE | Hotelling $t$ | $p$ |
|---|---|---|---|---|---|---|---|
| Intercept | $\alpha_0$ | 1.715 | 1.461 | 1.968 | 0.124 | 191.3 | <0.001 |
| Past 30-minute step count | $\alpha_1$ | 0.414 | 0.351 | 0.477 | 0.031 | 182.0 | <0.001 |
| At home/work | $\alpha_2$ | 0.143 | -0.083 | 0.368 | 0.110 | 1.67 | 0.205 |
| Walking Suggestion | $\beta_0$ | 0.050 | -0.167 | 0.267 | 0.106 | 0.22 | 0.640 |
| Walking Suggestion x At home/work | $\beta_1$ | 0.377 | 0.001 | 0.753 | 0.184 | 4.18 | 0.049 |
| Anti-sedentary Suggestion | $\beta_2$ | 0.092 | -0.166 | 0.351 | 0.127 | 0.53 | 0.472 |
| Anti-sedentary Suggestion x At home/work | $\beta_3$ | -0.142 | -0.540 | 0.256 | 0.195 | 0.53 | 0.472 |

*Note.* LCL (UCL) represents lower (upper) confidence limit. SE represents standard error. LCL, UCL, SE, and *p* are corrected for small sample size using method in (Liao et al., 2016; Mancl & DeRouen, 2001). The degrees-of-freedom for the Hotelling *t* test is (1, 30).

Table 4.

*Estimated effect of activity suggestion on proximal outcome, moderated by whether activity planning support was received on previous night*

| Variable | | Estimate | 95% LCL | 95% UCL | SE | Hotelling $t$ | $p$ |
|---|---|---|---|---|---|---|---|
| Intercept | $\alpha_0$ | 1.764 | 1.511 | 2.017 | 0.124 | 201.3 | <0.001 |
| Past 30-minute step count | $\alpha_1$ | 0.414 | 0.351 | 0.476 | 0.031 | 180.5 | <0.001 |
| Planning on previous day | $\alpha_2$ | 0.050 | -0.106 | 0.205 | 0.076 | 0.43 | 0.518 |
| Activity suggestion | $\beta_0$ | 0.113 | -0.035 | 0.261 | 0.073 | 2.43 | 0.129 |
| Activity suggestion x Planning on previous day | $\beta_1$ | 0.046 | -0.228 | 0.320 | 0.134 | 0.12 | 0.734 |

*Note.* LCL (UCL) represents lower (upper) confidence limit. SE represents standard error. LCL, UCL, SE, and *p* are corrected for small sample size using method in (Liao et al., 2016; Mancl & DeRouen, 2001). The degrees-of-freedom for the Hotelling *t* test is (1, 32).

Supplementary Figure 1.

*Estimated effect of activity suggestion on proximal outcome as a linear function of days in study, and corresponding 95% pointwise confidence intervals*

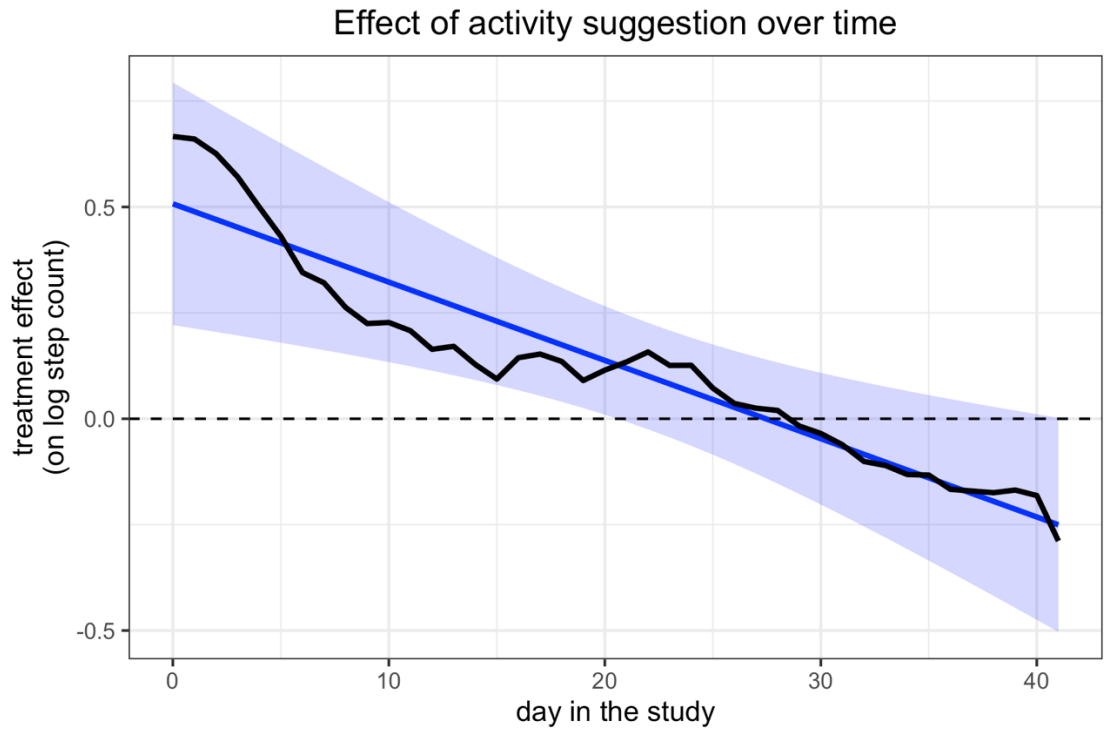

*Note.* Figure for the sensitivity analysis in Appendix D regarding "Question 2: Does the effect of the activity suggestions change with each additional day in the study?" in section "Analysis Using Data from HeartSteps MRT." The black curve is the estimated effect using local 2-degree polynomial regression with smoothing span 2/3 and tricubic weighting. The blue line represents the estimated causal excursion effect across the 42 study days, assuming a linear time trend, and the shaded blue area is the pointwise 95% confidence interval.

Supplementary Table 1.

*Simulation results for Appendix C: Efficiency gain from including prognostic variable in the working model*

|        | bias   | standard deviation | 95% coverage probability |
|--------|--------|--------------------|--------------------------|
| WCLS-1 | -0.001 | 0.067              | 96.7%                    |
| WCLS-2 | -0.001 | 0.067              | 96.9%                    |
| WCLS-3 | -0.001 | 0.074              | 95.8%                    |
| WCLS-4 | -0.001 | 0.074              | 95.7%                    |

*Note.* All four WCLS assumes the constant treatment effect model, and they differ in the choice of the working model. WCLS-1 includes control variables $(1, X_t, Y_t)$; WCLS-2 includes control variables $(1, X_t)$; WCLS-3 includes control variables $(1, Y_t)$; WCLS-4 includes only the intercept.